\documentclass[a4paper,twocolumn,english,prl]{revtex4-2}
\usepackage[T1]{fontenc}
\usepackage[latin1]{inputenc}
\usepackage{color}
\usepackage{textcomp}
\usepackage{amsmath}
\usepackage{amssymb}
\usepackage{graphicx}
\usepackage{subscript}
\usepackage{hyperref}

\makeatletter

\pdfpageheight\paperheight
\pdfpagewidth\paperwidth

\usepackage{babel}
\usepackage{amsfonts}

\makeatother

\usepackage{babel}
\begin{document}
\title{EEGs disclose significant brain activity correlated with synaptic
fickleness}
\author{Jorge Pretel}
\author{Joaquín J. Torres}
\email{Corresponding author: jtorres@onsager.ugr.es}
\author{J. Marro}
 \affiliation{Institute ``Carlos I'' for Theoretical and Computational Physics,
University of Granada, Spain}

\begin{abstract}
We here study a network of synaptic relations mingling excitatory
and inhibitory neuron nodes that displays oscillations quite similar
to electroencephalogram (EEG) brain waves, and identify abrupt variations
brought about by swift synaptic mediations. We thus conclude that
corresponding changes in EEG series surely come from the slowdown
of the activity in neuron populations due to synaptic restrictions.
The latter happens to generate an imbalance between excitation and
inhibition causing a quick explosive increase of excitatory activity,
which turns out to be a (first-order) transition among dynamic mental
phases. Besides, near this phase transition, our model system exhibits
waves with a strong component in the so-called \textit{delta-theta
domain} that coexist with fast oscillations. These findings provide
a simple explanation for the observed \textit{delta-gamma} and \textit{theta-gamma
modulation} in actual brains, and open a serious and versatile path
to understand deeply large amounts of apparently erratic, easily accessible
brain data. 
\end{abstract}
\maketitle
Today one successfully associates most brain activity with events
in which large sets of neurons cooperate arbitrated by willful variations
of their synaptic relations \cite{marrotorresbook}. This broadcasts
signals throughout, and EEG exploration on the cerebral cortex has
thus become a relatively simple, convenient and inexpensive way of
analyzing consequences of such an intriguing collaboration \cite{wright99,barras16,ruijter17,lopezmadrona20}.
In fact, EEG studies deliver some overall image of the brain activity
with good time accurateness that complements magnetic resonance analysis
of better spatial resolution. Specifically, EEGs watch over frequencies
and often distinguish $\delta,\theta,\alpha,\beta$ \ and $\gamma$\ ``rhythms''
---subsequently along the range 0.5 Hz to 35 Hz and more---, which
are loosely associated to different states of consciousness such as
say, deep sleep, anesthesia, coma, relax, and attention. 

Truly, this is at present a main noninvasive tool to deepen on the
brain operation under both normal and pathological conditions \cite{barras16,smith05,palva07,faigle13,houmani18},
and it is therefore convenient digging out on the interpretation of
all of those waves details. Indeed, a number of prototypes have already
addressed the origin and nature of observed brain oscillatory behavior,
e.g. \cite{wright99,lopes74,dafilis01,bojak04,liley09}. Recently,
following a hint \cite{lopes74} that $\alpha$ rhythms might come
out from filtering of cooperative signs by interactions with noisy
signals from different parts of the nervous system, it was explained
the emergence of a wide spectrum of brain waves within a simple computational
framework \cite{galadi20}. More specifically, this study has shown
that a neural module can exhibit waves in a variety of frequency bands
just by tuning the intensity of a noisy input signal. We interpret
this result as suggesting that a unifying mechanism in some way occurs
at some level of brain activity for a range of oscillations. In fact,
existing literature by now has described \cite{marrotorresbook} various
well defined, let us say, \textit{dynamic phases}, as well as transitions
among them ---typically, from states with a low and incoherent activity
to others that show a high synchrony--- where weak signals are processed
efficiently in spite of much unrests around. One thus understands,
for instance, that this ability is due to the very large susceptibility
developed in the medium by a phase transition due to a mechanism occasionally
termed as \textit{stochastic resonance }\cite{torres13}.

The picture in previous theoretically-oriented EEG work, including
\cite{galadi20}, is mostly phenomenological and generally adopts
a uniform and stationary description of the neuron relations efficiency,
thus forgetting the actual possibility that synapses perform dynamically
during the neuron cooperation processes \cite{tsodyks97,zucker02,tsodyks00,marrotorresbook}.
Nevertheless, synaptic relations surely vary with time while affecting
essentially both the neuron network global behavior and the ensuing
capacities to transmit information \cite{tsodyks00,abbott02,tsodyks98,mejias08,mongillo08,torreskappen13,marrotorresbook}.
For instance, a sort of sudden synaptic facilitation can allow for
transient persistent activity after removal of a stimulus \cite{mongillo08},
which may be the basis for working memory. Moreover, It was reported
that synaptic dynamics may induce in the human cortex bursting escorted
of asynchronous activity \cite{tsodyks00}, as well as instabilities
prompting transitions among attractors, which allow for effective
memory searching \cite{torreskappen13}, in addition to a kind of `up-and-down
states' reported to occur in cortical neuron populations \cite{holcman06}.
Aware of these and similar facts, we mathematically recast and generalize
here both the more mesoscopic description in \cite{wright99} and
the algorithmic model in \cite{galadi20}, perfecting them with detailed
dynamic synapses and other realistic features. We thus show how certain
levels of short-time 'depression' of the synaptic links induce transitions
between states of synchronized excitatory-inhibitory neuron populations
and global states of incoherent behavior. It follows that one may
speak of kind of sharp phase transitions, clearly displaying metastability
and hysteresis, that have been experimentally observed \cite{kim18}.
Furthermore, near such explosive variations, our model exhibits oscillations
with a prominent component in the $\delta-\theta$ band along with
high frequency activity, namely, the $\delta-\gamma$ and $\theta-\gamma$
modulations already perceived in actual brain EEG recordings \cite{lisman13,headley17},
which have been associated with \textit{fluid} intelligence \cite{gagol18}.
Even more, we here associate such intriguing sharp variations with
disruptions of the balance among excitation and inhibition produced
by depression of excitatory inputs into inhibitory neurons. This reduces
the inhibitory activity thus prompting a sudden excitation increase
that further reduces inhibition. Interestingly enough, a lack of the
excitation-inhibition balance in the actual human brain could be crucial
to understand the essentials of some recurrent neurological disorders
such as epilepsy, autism and schizophrenia, e.g., \cite{lee17,sohal19}.

The simplest version of our model aims to capture the essentials of
the cerebral cortex operation allowing for a network with excitatory
(E) and inhibitory (I) neurons, the former occurring four times the
latter. Furthermore, the amplitude of the corresponding postsynaptic
responses follows the opposite ratio, i.e., the response evoked by
any I is four times larger than that by any E. This is supposed to
correspond to a realistic \textit{cortex balanced state} \cite{meinecke87,heiss08}.
We then represent a region of the cerebral cortical tissue with a
large square of \textit{N} nodes with periodic boundary conditions
and fulfilling such a balance, in which each I node influences a set
of 12 neighboring E's and it is influenced by 32 adjacent E's. As
in previous work \cite{lopes74,galadi20}, we do not consider here E-E and I-I
feedbacks. Besides, from the various familiar types of imaginable
neuron dynamics, we refer to the celebrated integrate-and-fire case
\cite{burkitt06,marrotorresbook}. Namely, the cell membrane acts
as a capacitor subject to several currents, which results in a potential
\textit{V} for each neuron changing with time according to

\begin{equation}
\tau\frac{\mathit{dV}}{\mathit{dt}}=-V+V^{in}+V^{\mathit{noise}}.\label{eq:1}
\end{equation}
Here, as in previous work \cite{lopes74,galadi20}, the time constant
$\tau$ is set equal to \textit{$\tau_{1}$}\textsubscript{}\textit{
}(\textit{$\tau_{2}$}\textsubscript{}) according the membrane cell
voltage is above (below) certain resting potential, which we set to
zero. The last two terms in (1) correspond to the voltage induced
by the sum of all currents through the membrane, which we separate
here in two main contributions. $V^{in}$ is the sum of inputs from
adjacent neighbors that influence the given cell, while $V^{\mathit{noise}}$
accounts for any input from neurons in other regions of the brain.
Assuming lack of correlations \cite{shadlen98}, we represent $V^{\mathit{noise}}$
as a Poisson signal characterized by a noise level parameter \textit{$\mu$.}

It is now well established that, in human brains, synapses linking
neurons may undergo variations in scales from milliseconds to minutes,
in addition to more familiar long-term plastic effects. In fact, one
observes short-term depression (STD), in which the synaptic efficacy
decreases due to depletion of neurotransmitters inside the \textit{synaptic
button} after heavy presynaptic activity \cite{tsodyks97}. In addition,
there was reported kind of short-term facilitation characterized by
an increase of the efficacy strength \cite{henning13,bertram96,jackman17},
which results from a growth of the intracellular calcium concentration
after the opening of the voltage gated calcium channels due to successive
arrival of action potentials to the synaptic button. It seems that,
in general, these two short-term mechanisms may compete \cite{marrotorresbook,mejias08} but, for simplicity, we just consider here synapses endowed of STD,
and describe this by using the release probability\textit{ $U$} and
the fraction of neurotransmitters at time \textit{t} ready (to be
released) after the arrival of an action potential\textbf{ $x_{t}$} \cite{tsodyks98}. The ensuing image is that, each time a presynaptic
spike occurs, a constant portion \textit{U} of the resources \textbf{$x_{t}$
}is released into the synaptic gap, and the remaining fraction\textbf{
$1-x_{t}$ }becomes available again at rate\textbf{ $1/\tau_{\mathit{rec}}$}.
Therefore,

\begin{equation}
\frac{\mathit{dx}_{t}}{\mathit{dt}}=\frac{1-x_{t}}{\tau_{\mathit{rec}}}-Ux_{t}\delta\left(t-t_{\mathit{sp}}\right),\label{eq:2}
\end{equation}
where the delta function makes that the second right-hand term only
occurs for $t=t_{\mathit{sp}}$, the time at which a presynaptic input
spike arrives. Assuming also the amplitude of the response proportional
to the neurotransmitters fraction released after the input spike,
the STD effect can be written, for E and I neurons respectively, as
follows

\begin{equation}
V_{t}^{in,d}=V_{0}^{d}Ux_{t_{\mathit{sp}}}\left[\Theta\left(t-t_{\mathit{sp}}\right)-\Theta\left(t-t_{\mathit{sp}}-t_{\mathit{max}}\right)\right]\label{eq:3}
\end{equation}

\begin{equation}
V_{t}^{in,h}=V_{0}^{h}Ux_{t_{\mathit{sp}}}\Theta\left(t-t_{\mathit{sp}}\right)e^{\frac{-(t-t_{\mathit{sp}})}{\tau_{2}}}\label{eq:4}
\end{equation}
where $\Theta(X)$ \ is the Heaviside step function. The form of
these inputs generated by E and I neurons are chosen so that the response
generated on the postsynaptic neuron membrane matches data; see, for
instance, \cite{galadi20}. Thus, for simplicity, we model the excitatory
synaptic input by a square pulse of width $t_{\mathit{max}}$ and
maximal amplitude $V_{0}^{d}$\textcolor{black}{,} as described by
Eq. (3), and the inhibitory input by a decaying exponential behavior
with time constant $\tau_{2}$ \ and maximum amplitude $V_{0}^{h}$,
as in Eq. (4).\textbf{ }In addition, to account for synaptic strength
variations that depend on presynaptic history, we multiply these input
functions by a factor $U{\cdot}x_{t_{\mathit{sp}}}$\textit{, }thus
ensuring that the amplitude of the synaptic input is proportional
to the amount of neurotransmitters released right after a presynaptic
spike, which is an activity dependent factor through dynamics in Eq.
(2). Note that there is no synaptic variability present when $U{\cdot}x_{t_{\mathit{sp}}}=\mathit{constant}$
\ occurring for $\tau_{\mathit{rec}}\rightarrow0.$ \ \textcolor{black}{Furthermore,
to prevent the membrane potential in (1) from reaching physiologically
unrealistic levels, we impose upper and lower limits of } $V_{\mathit{sat}}=90\mathit{mV}$
\textit{\textcolor{black}{\ }}\textcolor{black}{and } $V_{\mathit{min}}=-20\mathit{mV}$\textcolor{black}{,
respectively, around the resting membrane potential, } $V_{\mathit{rest}}=0$
\textcolor{black}{\ as said. This is achieved by multiplying the
different excitatory and inhibitory inputs by the terms} $(V_{\mathit{sat}}-V)/V_{\mathit{sat}}$
\textcolor{black}{\ and} $(V_{\mathit{min}}-V)/V_{\mathit{min}}$\textcolor{black}{,
respectively. }

\textcolor{black}{Equations (1)-(4) fully describe the dynamics of
the membrane potential in our basic model below a threshold for firing,
which is in principle set } $V_{\mathit{th}}=6\mathit{mV}$ \ above
the resting membrane potential for both E and I neurons. Additionally,
after generation of a spike at $t_{f}$, we assume an absolute refractory
period ($t_{a}$) during which the neuron is unable to fire again,
and a subsequent relative refractory in which the ability to produce
new spikes is constrained. Therefore, we set 

\[
V_{\mathit{th}}(t)=\left\{ \begin{matrix}V_{\mathit{sat}}t_{f}<t<t_{f}+t_{a}\\
6+\left(V_{\mathit{sat}}-6\right)e^{-\kappa\left(t-t_{f}-t_{a}\right)}t>t_{f}+t_{a}.
\end{matrix}\right.
\]
That is, the threshold is first set to $V_{\mathit{sat}}$ (during
one hundred time steps, which gives $t_{a}=4\mathit{ms}$) to impede
any further spike generation during $t_{a}$. Then, it decays exponentially
to its resting value of $6\mathit{mV}$ with a time constant $\kappa^{-1}=0.5\mathit{ms}$
\ that mimics the existence of a relative refractory period.

{ Using this clear-cut and supposedly realistic model, we numerically
analyzed how synaptic STD affects eventually emergent waves by carefully
monitoring the network dynamics for adiabatically increasing values
of the noise parameter \textit{{\textmu }}. Figure 1 depicts the
resulting average membrane potential of the E population versus \textit{{\textmu }},
which clearly illustrates the mentioned sharp transitions. That is,
in the absence of STD (top panel in each column), waves do not vary
essentially with the external noise amplitude within \textit{{\textmu }
}${\in}$ (0.5, 100), as already reported in \cite{galadi20}. This
regime corresponds to the simplest and most familiar brain waves.
However, when STD is on {}---namely, the synaptic efficacies vary
with the system activity so that parameter $\tau_{\mathit{rec}}$
is large enough{}--- an `explosive' transition may show up as \textit{{\textmu }}
increases. This occurs at lower $\tau_{\mathit{rec}}$ \ the lower
the maximal excitatory postsynaptic amplitude $V_{0}^{d}$ \ is.
It is said `explosive' in the sense that the transition shows hysteresis,
from well-defined synchronized behavior to a state of high excitation
and low coherence, as we vary \textit{{\textmu }} adiabatically
forward (purple line) and backward (green line). Ona may also name
this a \textit{first-order phase transition} by simple analogy with
thermodynamics, though with the warning that the present setting is
a nonequilibrium one \cite{marro05}. }

\begin{figure}
\begin{centering}
\includegraphics[width=8.5cm]{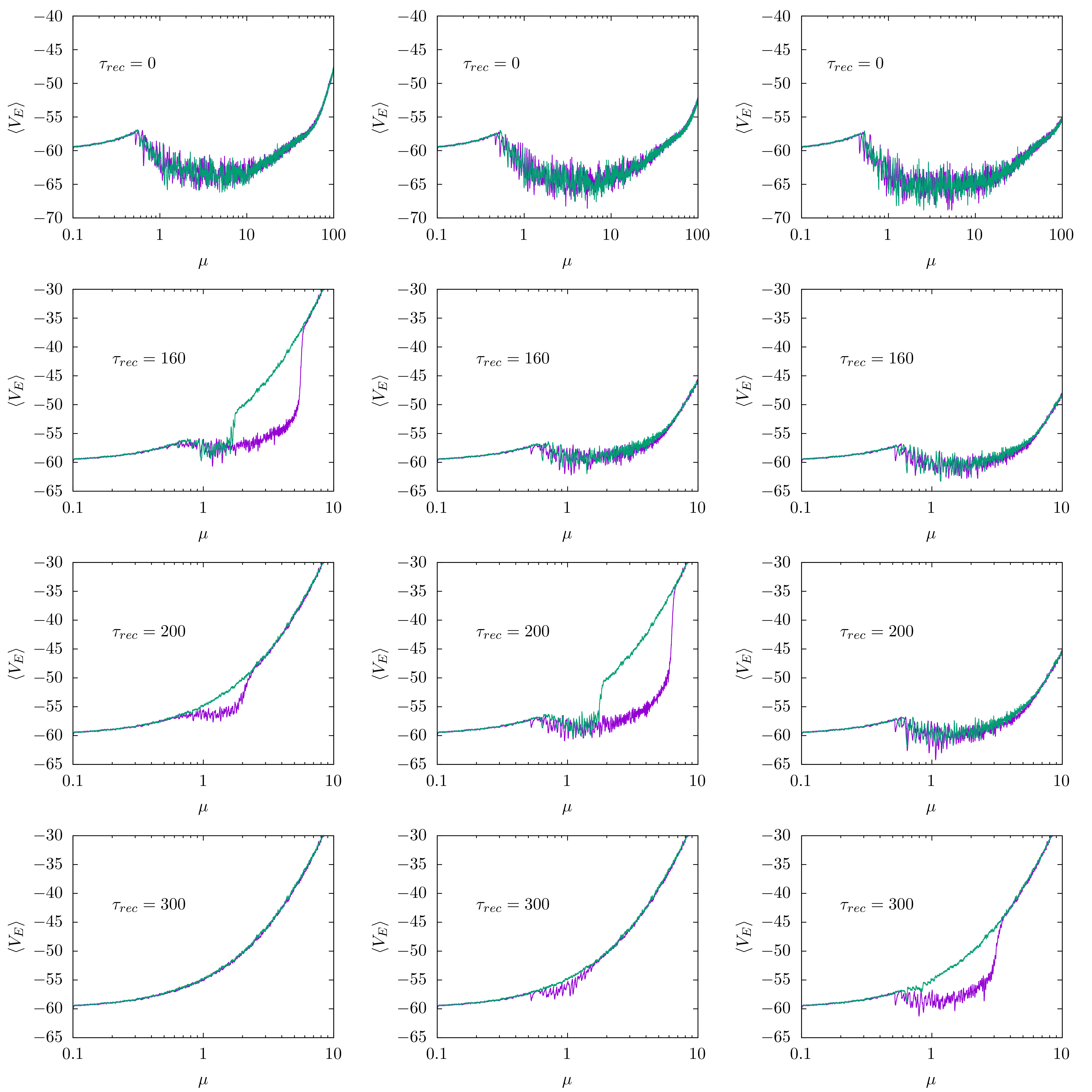} 
\par\end{centering}
\caption{Evidence for sharp changes in emergent cooperative-neuron EEG-like
waves as the noise level \textit{{\textmu }} varies when synaptic
depression is set on. Columns are, from left to right, for $V_{0}^{d}=8,10$
\ and $12\mathit{mV}$, respectively and $V_{0}^{h}=-4V_{0}^{d}$.
In all cases, $U=0.5,\tau_{1}=16\mathit{ms},$ $\tau_{2}=26\mathit{ms}$
\ and external excitatory noisy inputs modelled inducing each one
a constant depolarization $V_{o}^{d}=5.48\mathit{mV}$. This is for
a module with 196 E's and 49 I's. }
\end{figure}
The resulting phase diagram in the $\left(\mu,\tau_{\mathit{rec}}\right)$
space is illustrated in figure 2. The solid quasi-vertical line, for
$\mu<0.5,$ \ describes a (continuous or second-order) phase transition
between a near silent phase A, with asynchronous sporadic spikes at
low rate (corresponding to the asynchronous down state actually observed
in the brain), and an oscillatory phase B, where brain waves emerge
with increasing frequency as $\mu$ \ increases (see figure 1). As
$\tau_{\mathit{rec}}$ \ increases in the system, figure 2 indicates
that the brain waves disappear at a (first-order) transition (dashed
line), where a new phase D of waves with high excitation and low coherence
emerges. This sharp transition becomes smooth above a say `tricritical
point' (1.4, 268) (short quasi-vertical solid line on top). The small
region C shows metastability as revealed by hysteresis. Note that
when $\mu$ is large this region C narrows as noise level increases.
In addition, region B contains (red and blue) areas in which brain
waves sharply emerge with high values of the firing rates ({\textgreater}100
Hz) for E and I neurons.

\begin{figure}
\begin{centering}
\includegraphics[width=8.5cm]{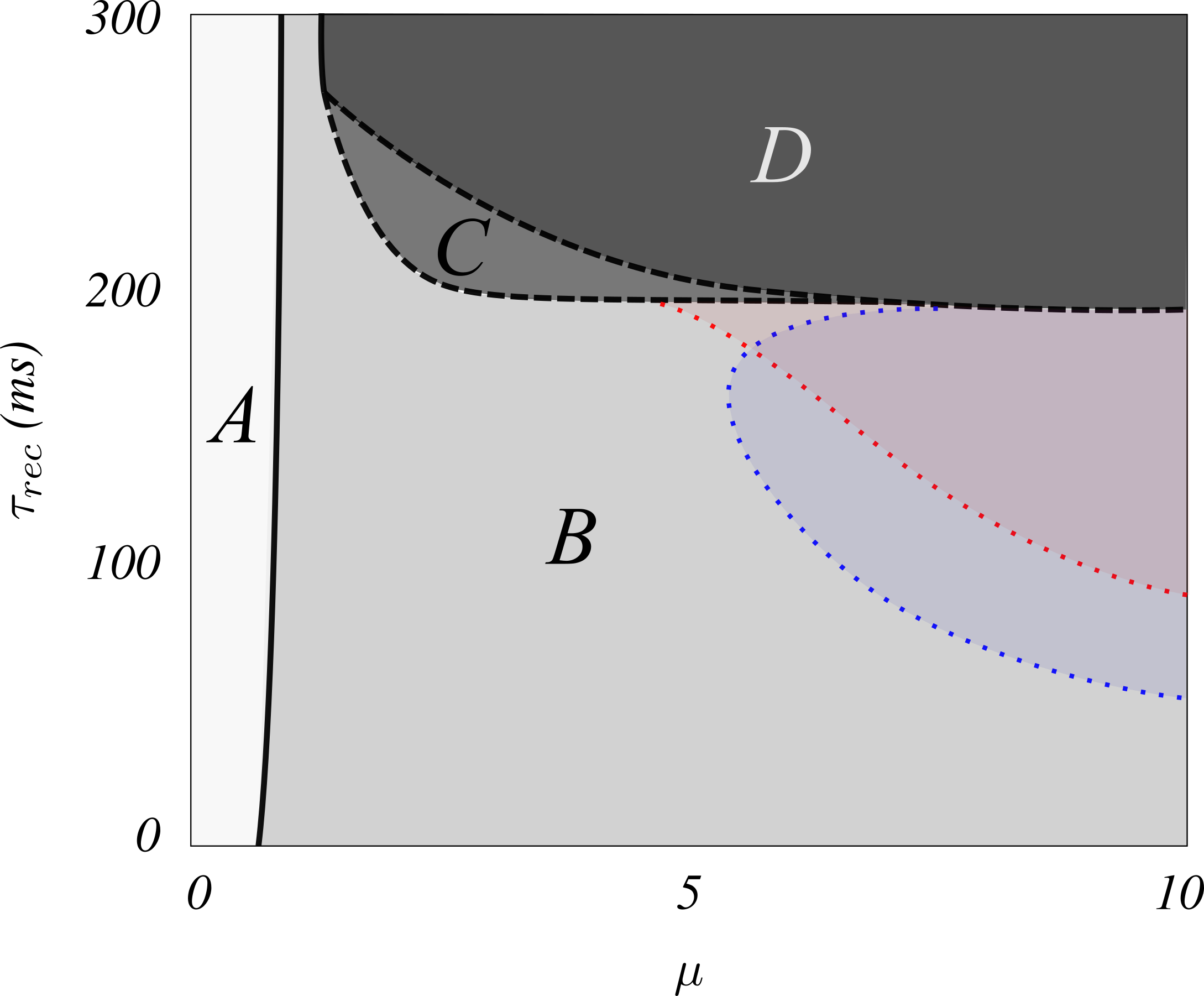} 
\par\end{centering}
\caption{Diagram $(\mu,\tau_{\mathit{rec}})$ illustrating different (dynamical)
phases in our system. For low noise (region A), there are random sporadic
excitatory firing events unable to depolarize the I neurons. Region
B shows well-defined rhythms ranging from $\alpha$ \ to $\gamma$
\ bands, while a higher depression induces ceasing of the inhibitory
activity and a consequent absence of synchronicity and coherence in
region D. Metastability as in figure 1 characterizes the region C.
Red and blue colored areas in B indicate emerging waves with high
values of the firing rates ({\textgreater}100 Hz) for E and I neurons,
respectively. Dashed lines illustrate first-order phase transitions,
while continuous lines denote second-order transitions.}
\end{figure}
Trying to deep on the nature of the sharp transition, we monitored
(figure 3) the change with depression of both the mean firing rate
and the mean amplitude of the oscillations in E and I neuron populations
when it occurs (for \textit{{\textmu } }= 3). This shows that, as
STD increases, E neurons induce the I ones to slowly decaying their
firing rates as approaching the transition point, where they become
silent. A feedback induced by this decay of the I activity makes the
E's to increase their firing activity until reaching (at the transition
point) its maximum possible, then remaining firing at the maximum
possible frequency. This induces important facts on the ensuing oscillations:
the amplitude of the inhibitory component of the waves jumps to zero
at the transition point, and the amplitude of the excitatory component
decays to a very low value below $V_{\mathit{th}}.$

\begin{figure}
\begin{centering}
\includegraphics[width=8.5cm]{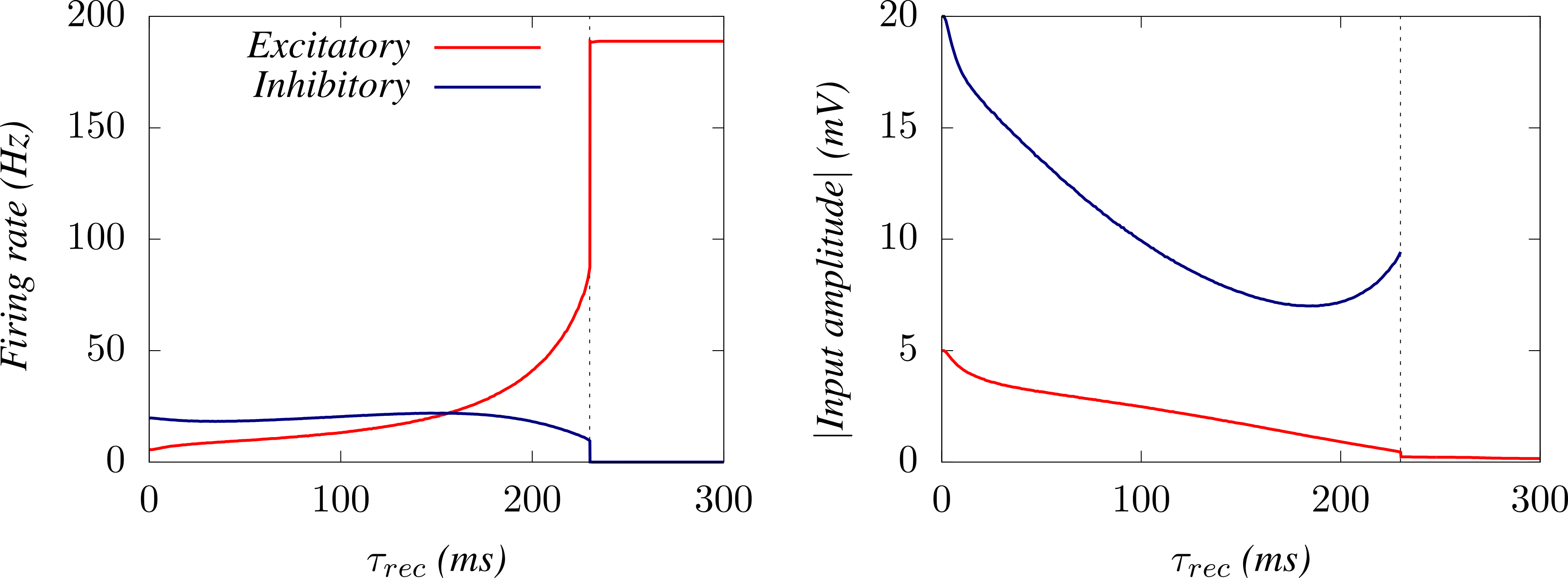} 
\par\end{centering}
\caption{\textbf{Left}: Average firing rate for E and I neurons as the level
of STD is increased until the explosive transition occurs for an external
depolarizing noise\textit{ $\mu=3$}. \textbf{Right}: Corresponding
average amplitude of the oscillations. This illustrates that the transition
occurs because of a cease of firing of I's due to the negative feedback
of highly depressed I's over E's, which then star to fire at the transition
point at the higher frequency, thus depressing even more the I's until
impeding their firing.}
\end{figure}
Also interesting is how the nature of the emerging waves changes with
STD. For a relatively low noise, e.g. \textit{{\textmu }} = 0.8,
the network's response remains nearly unchanged, while the amplitude
of the oscillations decays until no well-defined oscillatory behavior
is observed as STD is increased (figure 4, case \textit{{\textmu }}
= 1). Note that this transition from a state with synchrony to an
incoherent one become abrupt as described above for a level of noise
$\mu>1$\ (see figure 5). 

\begin{figure}
\begin{centering}
\includegraphics[width=8.5cm]{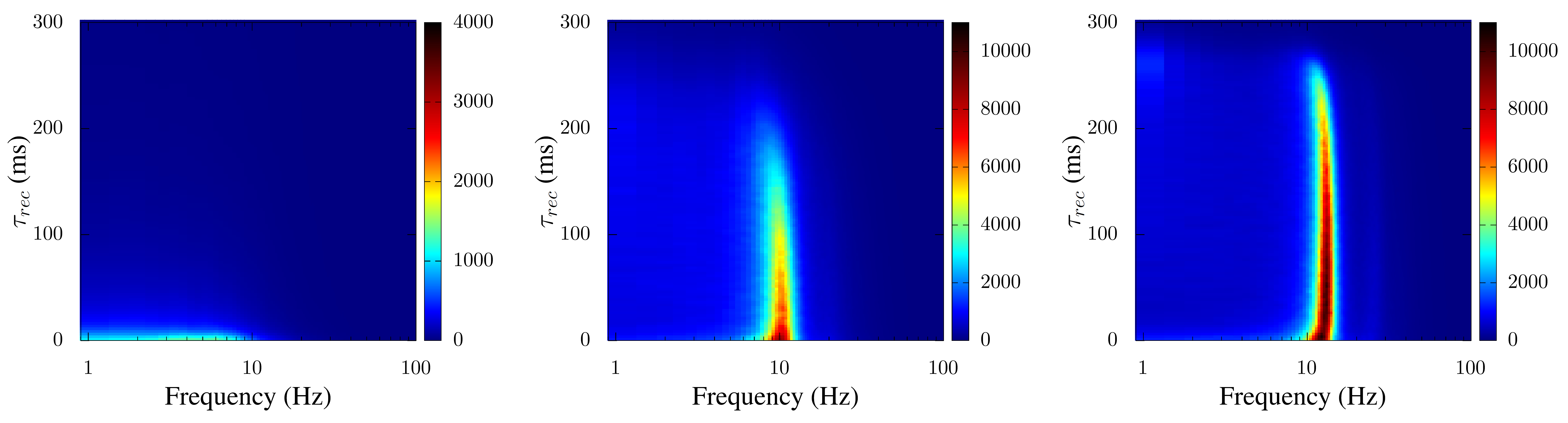} 
\par\end{centering}
\caption{Emergence of ``$\alpha$ rhythms'' (around 10Hz) in the model for
noise levels \textit{{\textmu }} = 0.6, 0.8 and 1.0, respectively,
from left to right. Although the power of the main frequency of the
waves decays as STD increases, this illustrates how these waves details
are not dramatically affected by synaptic depression and the $\alpha$
\ band regime remains until $\tau_{\mathit{rec}}{\approx}260\mathit{ms}$,
where the waves disappear (note that this transition becomes explosive
for $\mu{\gtrsim}1$ \ as shown in figure 5). }
\end{figure}
For higher values of \textit{{\textmu } }(e.g., $\mu=3$ in figure
5), the power spectrum of the response shows significant changes.
First, its peak frequency notably increases for higher levels of STD,
becoming up to twice as great as for the static case ( $\tau_{\mathit{rec}}=0$),
thus inducing waves in the $\beta$ \ and $\gamma$ \ regimes. This
STD-induced transition from low to high frequency bands confirms that
synaptic plasticity could be an important mechanism in modulating
the nature of the oscillations from cortical neuronal populations.
In addition, we observe that an increase of $\tau_{\mathit{rec}}$
\ can produce secondary, low-frequency peaks coexisting with the
main peak in the power spectrum of the emergent waves. This phenomenon
is most evident near the explosive transition point, where a prominent
component in the $\delta/\theta$ bands emerges, accompanied by a
general enhancement in the amplitude of the oscillations, as can be
seen in the time series presented in figure 5. This effect seems to
occur for all relatively high levels of noisy, namely, $\mu>1.$ 
\begin{figure}
\begin{centering}
{\includegraphics[width=8.5cm]{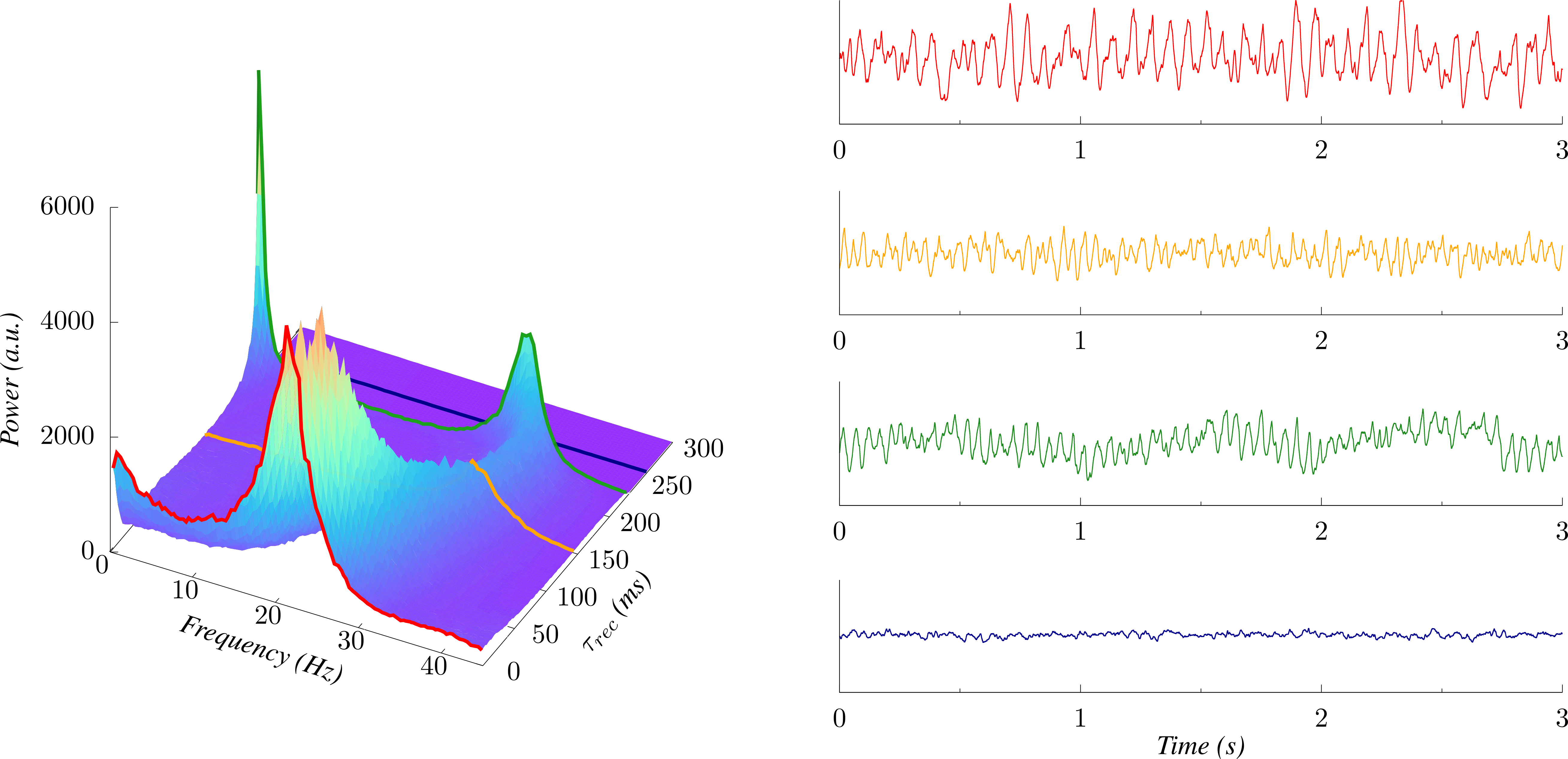} } 
\par\end{centering}
\caption{Left: Power spectra of the system response as a function of the recovery
time \textit{$\tau_{rec}$} for $\mu=3$. Right: time series of the
emergent oscillations for particular levels of synaptic depression,
namely, \textit{$\tau_{rec}$} = 0, 145, 230 and 265 ms, respectively,
from top to bottom. The associated power spectra for each of these
series are highlighted (with the same color) in the surface plot of
the left panel.}
\end{figure}

\begin{figure}
\begin{centering}
\includegraphics[width=8.5cm]{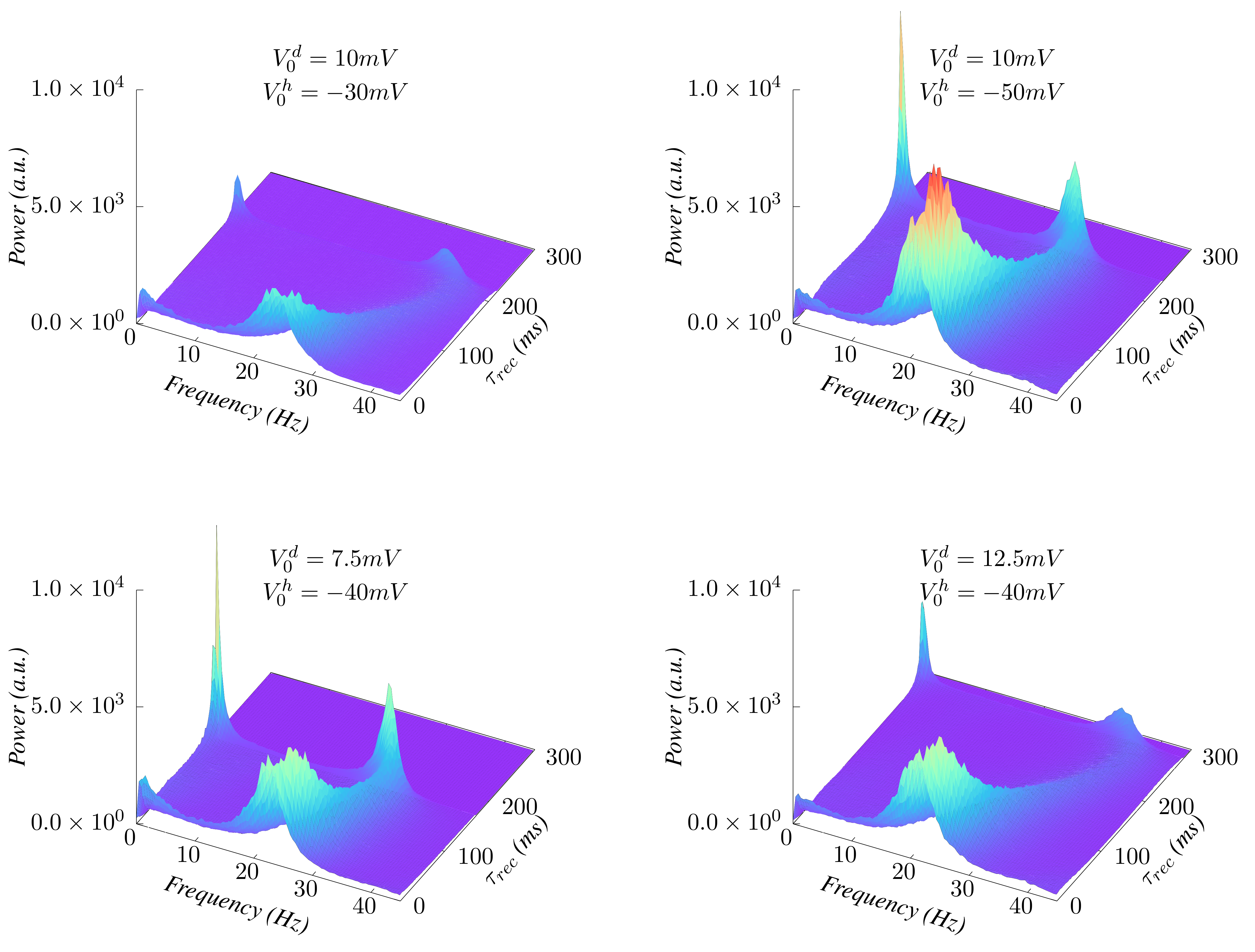} 
\par\end{centering}
\caption{Effect of varying the E/I amplitude ratio as depression increases.
\textbf{Top:} Increasing the amplitude of I's while leaving the E's
unchanged enlarges the $\delta/\theta$ \ component of the $\delta(\theta)-\gamma$
\ modulation around the phase transition ( $\tau_{\mathit{rec}}{\approx}230\mathit{ms}$).
\textbf{Bottom: }Increasing the E's while maintaining the I's moves
the transition to higher levels of depression and makes the emergent
oscillations more sensitive to synaptic depression.}
\end{figure}
Concerning the effect of the E/I balance on emergent behavior, it
interests how it affects the incidence of $\delta\left(\theta\right)-\gamma$
\ modulations around the transition, and how the appearance of this
is affected by the level of synaptic depression. Figure 6 illustrates
some effects of the ratio between the E and I synaptic efficacies.
We observe that, when $V_{0}^{d}/V_{0}^{h}$ \ decreases and the
inhibitory synapses become relatively more influential, the low frequency
$\delta/\theta$ \ component becomes more significant for oscillatory
behavior (see Figure 6, top-right and bottom-left panels) while, when
this ratio increases, the low frequency band components ($\delta$
and $\theta$) tend to disappear (Figure 6, top-left and bottom-right
panels). Additionally, an increase of $V_{0}^{d}/V_{0}^{h}$, which
implies more excitation, makes the oscillations frequency more susceptible
to changes on synaptic depression (see Figure 6, top-left and bottom-right
panels), while a stronger inhibition tends to maintain the frequency
of the emergent waves nearly unchanged against depression increases
(Figure 6, top-right and bottom-left panels).

Summing up, we present in this Letter a very simple model that, recasting
previous EEG related work, has two significant features. One is that
it provides a well-defined set-up to undertake a systematic interpretation
of apparently erratic brain EEG data. These are easily accessible
today and, as we have foreseen here, happens to carry important information
concerning the brains activity. Furthermore, this model is convenient
to admit complements that one might suspect to be relevant in these
scenarios such as, for instance, other synaptic mechanisms, complex
synaptic networks and more realistic node neurons. In addition, and
perhaps even more transcendental within this context, the framework
presented here precisely illustrates how the concept of a (nonequilibrium)
phase transition \cite{marro05} may be essential for an accurate
description of the brain properties.

\begin{acknowledgements}
The authors acknowledge financial support from the Spanish Ministry
of Science and Technology, and the Agencia Española de Investigación
(AEI) under grant FIS2017-84256-P (FEDER funds) and from the Consejería de Conocimiento, Investigación Universidad, Junta de Andalucía and European Regional
Development Fund, Ref. A-FQM-175-UGR18. 
\end{acknowledgements}

\end{document}